\documentclass[preprint,12pt]{elsarticle}  
\usepackage{graphicx} 
\usepackage{amssymb} 
\usepackage{amsmath} 

\journal{Physics Letters B} 

\begin{document}

\begin{frontmatter}

\title{Are there any narrow $K^-$- nuclear states?}

\author{Jaroslava Hrt\'{a}nkov\'{a}\corref{cor1}}
\ead{hrtankova@ujf.cas.cz}
\author{Ji\v{r}\'{\i} Mare\v{s}}
\cortext[cor1]{Corresponding author}%: Jaroslava Hrt\'{a}nkov\'{a}, hrtankova@ujf.cas.cz} 

\address{Nuclear Physics Institute, 250 68 \v{R}e\v{z}, Czech Republic}

\begin{abstract}
We performed self-consistent calculations of $K^-$-nuclear quasi-bound
states using a single-nucleon $K^-$ optical potential derived from chiral
meson-baryon coupled-channel interaction models, supplemented by a
phenomenological $K^-$ multinucleon potential introduced recently to
achieve good fits to kaonic atom data~\cite{fgNPA16}. Our calculations
show that the effect of $K^-$ multinucleon interactions on $K^-$
%binding energies and 
widths in nuclei is decisive. The resulting widths are considerably larger 
than corresponding binding energies. Moreover, 
when the density dependence of the 
$K^-$-multinucleon interactions derived in the fits of kaonic atoms is extended to the
nuclear interior, 
the only two models acceptable after imposing as
additional constraint the single-nucleon fraction of $K^-$ absorption at
rest do not yield any kaonic nuclear bound state in
majority of considered nuclei.
   
\end{abstract}


\end{frontmatter}

\section{Introduction}
\label{intro}
Interaction between the $K^-$ meson and nucleon(s) has been object of increased interest in recent years~\cite{NPA12,HYP12,EXA14,NPA16,HYP15}.  
The $K^- N$ interaction is closely related to such issues as the nature of the $\Lambda(1405)$ resonance, 
propagation of the antikaon in nuclear matter, production of strangeness or existence of $K^-$-nuclear 
quasi-bound states. Despite much effort in the last decade~\cite{W10,ghm16,s17}, %\cite{YA02, SGM07, IS07, DHW08, WH08, W10, BGL12, SR14, F05, D10, N16}
 the question of kaonic nuclear states is still not resolved. 

The $K^- N$ interaction has recently been described in the framework of chirally motivated meson-baryon 
interaction models, parameters of which have been tuned to fit low-energy experimental data. As was shown in Ref.~\cite{fgNPA16}, commonly accepted models provide quite diverse $K^- N$ scattering amplitudes below threshold. However, the amplitudes from this very energy region enter the construction of a 
$K^-$ -nucleus potential relevant for calculations of kaonic nuclear states. 

A distinctive feature of the $K^- p$ amplitudes is their strong energy dependence originating 
from the presence of the $\Lambda(1405)$  resonance, which is generated dynamically in the 
chiral coupled-channel models of meson-baryon interactions. It is thus imperative to treat 
the energy dependence of scattering amplitudes properly and evaluate the $K^-$-nucleus optical potential 
self-consistently~\cite{cfggmPLB, cfggmPRC11}. 

The chiral meson-baryon interaction models discussed in this work, Prague (P) \cite{pnlo} and Kyoto-Munich (KM) \cite{kmnlo}, include only $K^-$ absorption on a 
single-nucleon, $K^-N\rightarrow \pi Y$ ($Y= \Lambda, \Sigma$). Calculations of $K^-$-nuclear bound states 
 based solely on these chiral models yield $K^-$ absorption widths quite small due to the proximity 
of $\pi \Sigma$ threshold~\cite{cfggmPLB, cfggmPRC11, gmNPA}. However, in the nuclear medium  
$K^-$ multinucleon interactions take place as well~\cite{fgNPA16, fgb93, fgNPA} and their role increases with density and 
$K^-$ binding energy. Therefore, the $K^-$ multinucleon processes, absorption in particular, 
have to be taken into account in any realistic assessment of the $K^-$ widths 
(to lesser extent also $K^-$ binding energies) in the nuclear medium. In Refs.~\cite{cfggmPLB, cfggmPRC11, gmNPA}, the $K^-NN$ absorption was incorporated 
using a phenomenological potential fitted to kaonic atom data since the applied chiral model did not address such 
processes. Recently Sekihara et al.~\cite{sjPRC12} described the non-mesonic $K^-$ interaction channels within 
a chiral unitary approach for the $s$-wave ${\bar K}N$ amplitude and evaluated the ratio of mesonic to 
non-mesonic $K^-$ absorption at rest inside the medium. 
The experimental information about this ratio comes from bubble chamber experiments \cite{bubble1, bubble2, bubble3}.  Friedman and Gal performed fits of  
kaonic atom data for several recent chirally motivated meson-baryon coupled-channel interaction models \cite{fgNPA16}. Subsequent comparison with the single-nucleon fractions of $K^-$ absorption at rest provided strict constraint on the meson-baryon interaction models describing the single-nucleon $K^-$ potential as well as on the corresponding phenomenological $K^-$ multinucleon optical potentials. Only the P and KM models were found acceptable by this analysis.

In this work, we apply the above two interaction models in calculations of $K^-$- nuclear quasi bound states. The single-nucleon $K^-$ potential is supplemented by a 
corresponding phenomenological optical potential which describes $K^-$ multinucleon interactions. 
We demonstrate that the $K^-$ multinucleon interactions in the nuclear medium affect crucially the 
$K^-$ widths. For the first time, we perform calculations of kaonic nuclear states using 
$K^-$-nuclear potentials containing both $K^-$ single-nucleon and multinucleon interactions, while   
the multinucleon potential was fitted for each chiral $K^-N$ amplitude model to kaonic 
atom data separately and further confronted with branching rations of $K^-$ absorption at rest.

\section{Methodology}
\label{sec-1}
The self-consistent calculations of $K^-$-nuclear quasi bound states are based on solving the Klein-Gordon equation for $K^-$ in the medium
\begin{equation}\label{KG}
 \left[ \vec{\nabla}^2  + \tilde{\omega}_{K^-}^2 -m_{K^-}^2 -\Pi_{K^-}(\omega_{K^-},\rho) \right]\phi_{K^-} = 0~,
\end{equation}
which yields kaon binding energies $B_{K^-}$ and widths $\Gamma_{K^-}$. Here, $m_{K^-}$ denotes the $K^-$ mass, $\tilde{\omega}_{K^-} =m_{K^-} - B_{K^-} -{\rm i}\Gamma_{K^-}/2 -V_C= \omega_{K^-} - V_C$, $V_C$ is
the Coulomb potential, and $\rho$ is the nuclear density distribution. In the present work, the energy- and density-dependent kaon self-energy operator 
$\Pi_{K^-}$ is constructed from scattering amplitudes derived within chiral SU(3) meson-baryon coupled-channel interaction models: Prague (P) \cite{pnlo} and Kyoto-Munich (KM) \cite{kmnlo}. These models capture physics of the $\Lambda$(1405) resonance and reproduce low energy $K^-N$ observables, consisting of cross-sections for low-energy $K^- p$ processes (listed in ref.~\cite{kmnlo}), three accurately determined threshold branching ratios \cite{mNPB}, as well as the $1s$ level shift and width in the $K^-$ hydrogen atom from the SIDDHARTA experiment \cite{sidhharta}.     

The self-energy operator $\Pi_{K^-}$ entering Eq.~\eqref{KG} is constructed in a ``$t\rho$'' form as follows:
\begin{equation}\label{piK}
\Pi_{K^-} = 2\text{Re}( {\omega}_{K^-})V_{K^-}^{(1)}=-4\pi \frac{\sqrt{s}}{m_N}\left(F_0\frac{1}{2}\rho_p + F_1\left(\frac{1}{2}\rho_p+\rho_n\right)\right)~.
\end{equation}
Here, $F_0$ and $F_1$ denote the isospin 0 and 1 $s$-wave in-medium amplitudes, respectively, $m_N$ is the nucleon mass, $\sqrt{s}$ is the $K^-N$ total energy, and $V_{K^-}^{(1)}$ denotes the (single-nucleon) $K^-$-nucleus optical potential corresponding to the $K^- N$ amplitudes. The kinematical factor $\sqrt{s}/m_N$ transforms the scattering amplitudes from the two-body frame to the $K^-$-nucleus frame. The proton and neutron density distributions $\rho_p$ and $\rho_n$ in a given core nucleus are obtained within 
the relativistic mean-field model NL-SH~\cite{nlsh}. 

The in-medium amplitudes $F_0$ and $F_1$ are derived from the free-space amplitudes, $F_{K^-n}(\sqrt{s})$ and $F_{K^-p}(\sqrt{s})$, using the multiple scattering approach (WRW)~\cite{wrw} which accounts for Pauli correlations in the nuclear medium:
\begin{equation}\label{inmedAmp}
F_{1}=\frac{F_{K^-n}(\sqrt{s})}{1+\frac{1}{4}\xi_k \frac{\sqrt{s}}{m_N} F_{K^-n}(\sqrt{s}) \rho}~, \quad F_{0}=\frac{[2F_{K^-p}(\sqrt{s})-F_{K^-n}(\sqrt{s})]}{1+\frac{1}{4}\xi_k \frac{\sqrt{s}}{m_N}[2F_{K^-p}(\sqrt{s}) - F_{K^-n}(\sqrt{s})] \rho}~,
\end{equation}
where $\xi_k$ is adopted from Ref.~\cite{fgNPA16}.

The distinctive feature of $K^-p$ amplitudes constructed in chirally motivated coupled-channel models is their strong energy (and density) dependence  near and below threshold due to dynamically generated subthreshold $s$-wave resonance $\Lambda(1405)$. The energy dependence of in-medium scattering 
amplitudes calls for a proper self-consistent evaluation of corresponding $K^-$ optical potentials used in genuine calculations of $K^-$ atomic as well as nuclear states, as shown in Refs.~\cite{fgNPA16,cfggmPLB,gmNPA,fgNPA}.\\ 
The energy argument of in-medium amplitudes entering Eq.~\eqref{inmedAmp} is defined by Mandelstam variable 
\begin{equation}\label{mandelstam}
 s=(E_N+E_{K^-})^2-(\vec{p}_N+\vec{p}_{K^-})^2~,
\end{equation}
where $E_N=m_N-B_N$, $E_{K^-}=m_{K^-}-B_{K^-}$ and $\vec{p}_{N(K^-)}$ is the nucleon (kaon) momentum. The momentum dependent term in Eq.~\eqref{mandelstam} is no longer zero in the $K^-$-nucleus cm frame and generates additional downward energy 
shift~\cite{cfggmPLB}. The $K^-N$ amplitudes are expressed as a function of energy $ \sqrt{s} =~E_{th} + \delta \sqrt{s}$ where the energy shift $\delta \sqrt{s}$ can be approximated as
\begin{equation} \label{Eq.:deltaEs}
 \delta \sqrt{s}=  -B_N\frac{\rho}{\bar{\rho}}\, - \beta_N\! \left[B_{K^-}\frac{\rho}{\rho_{\rm max}} + T_N\left(\frac{\rho}{\bar{\rho}}\right)^{2/3}\!\!\!\! +V_C\left(\frac{\rho}{\rho_{\rm max}}\right)^{1/3}\right] + \beta_{K^-} {\rm Re}V_{K^-}(r)~.
\end{equation}
Here, $B_N=8.5$~MeV is the average binding energy per nucleon, $\bar{\rho}$ is the average nuclear density, $\rho_{\rm max}$ is the maximal value of the nuclear density, and $\beta_{N(K^-)}={m_{N(K^-)}}/(m_N+m_{K^-})$. $T_N=23$~MeV is the average nucleon kinetic energy in Fermi Gas Model. The energy shift respects the low-density limit, i.~e. $\delta \sqrt{s} \rightarrow 0$ as $\rho \rightarrow 0$ and the minimal substitution requirement $E \rightarrow E - V_C$ \cite{kkwPRL90}. Self-consistency is ensured by dependence of  $\delta \sqrt{s}$ on $B_{K^-}$, as well as on the $K^-$ optical potential $V_{K^-}$ determined by the energy dependent $K^-N$ in-medium amplitudes.
 
The $K^-$ interactions with two and more nucleons are an indispensable component of a $K^-$-nucleus interaction \cite{bfg97, mfgPLB, fgmNPA}. Recent analyses by Friedman and Gal have confirmed that the optical potential constructed from in-medium chirally motivated $K^-N$ amplitudes have to be supplemented by a phenomenological term representing $K^-$ multinucleon processes in order to achieve good fit to kaonic atom data \cite{fgNPA16, fgNPA}. Therefore, we supplement the single-nucleon potential $V_{K^-}^{(1)}$ by a phenomenological potential $V_{K^-}^{(2)}$ of the form
\begin{equation} \label{Vknn}
 2\text{Re}(\omega_{K^-})V_{K^-}^{(2)}=-4 \pi B (\frac{\rho}{\rho_0})^{\alpha} \rho~,
\end{equation}
where $B$ is a complex amplitude and $\alpha$ is a positive number. The parameters of the potential were fitted to kaonic atom data for both P and KM chiral meson-baryon 
interaction models separately. It has been shown in Ref.~\cite{fgNPA16} that only these two models are capable to reproduce simultaneously kaonic atom 
data and $K^-$ single-nucleon absorption fractions determined in bubble chamber experiments \cite{bubble1, bubble2, bubble3}. The corresponding values of the parameters $\alpha$, Re$B$ and Im$B$ including uncertainties are listed in Table~\ref{Tab.:ampB}. In view of the uncertainties (noticeably larger for $\alpha=2$), the P and KM models could be regarded as equivalent.

The full $K^-$ optical potential  $V_{K^-}$  used in a self-consistent evaluation of the subthreshold energy shift $\delta\sqrt{s}$ and in calculations of 
kaonic nuclear states is then constructed as a sum of the single- and multinucleon optical potentials $V_{K^-}=V_{K^-}^{(1)}+V_{K^-}^{(2)}$. 

\begin{table}[b!]
\begin{center}
\caption{Values of the complex amplitude $B$ and exponent $\alpha$ used to evaluate $V_{K^-}^{(2)}$  for chiral 
meson-baryon interaction models considered in this work.}\vspace{3pt}
 \begin{tabular}{c|cccccc|cc}
  & P1  & KM1 & P2 & KM2  \\ \hline %\hline
$\alpha$   & 1 & 1 & 2 & 2\\ 
Re$B$ (fm) &  $-$1.3 $\pm$ 0.2 & $-$0.9 $\pm$ 0.2 & $-$0.5 $\pm$ 0.6 & 0.3 $\pm$ 0.7 \\ 
Im$B$ (fm) &  ~~1.5 $\pm$ 0.2 & ~~1.4 $\pm$ 0.2 & ~~
4.6 $\pm$ 0.7 & 3.8 $\pm$ 0.7
 \end{tabular}
 \label{Tab.:ampB}
\end{center}
\end{table}  

In our calculations, we consider the conversion of $K^-$ on two nucleons $K^-NN \rightarrow \Sigma N$ to be the dominant mode of $K^-$ absorption in the nuclear interior \cite{sjPRC12, mfgPLB, sjPRC79}. The amplitudes Im$B$ for multinucleon processes are multiplied by a suppression factor which reflects the phase space reduction for decay products in $K^-NN\rightarrow \Sigma N$ absorption in 
the nuclear medium \cite{mfgPLB}. 

Experiments with kaonic atoms probe the real part of the $K^-$ optical potential reliably only up to $\sim 25$\% of $\rho_0$ and the imaginary part, that is dominant in causing strong interaction effects in kaonic atoms, is determined up to $\sim 50$\% of $\rho_0$ \cite{fgNPA16, bfg97}. Further in the nuclear interior, the shape of the phenomenological $K^-$ optical potential $V_{K^-}^{(2)}$ is mere extrapolation or analytic 
continuation of the empirical formula applied in the kaonic atom fit. 
Moreover, the larger value of the exponent $\alpha$ in Eq.~\eqref{Vknn}, the larger is sensitivity of extrapolations to the nuclear interior. 
Therefore, we consider two limiting options for $V_{K^-}^{(2)}$  
beyond the half density limit $\rho(r) = 0.5 \rho_0$ in our calculations. First, we apply the form~\eqref{Vknn} in the entire nucleus 
(full density option - FD). Second, we fix the potential $V_{K^-}^{(2)}$ at constant value $V_{K^-}^{(2)}(0.5\rho_0)$  
for $\rho (r) \ge 0.5~\rho_0$ (half density limit - HD). 

\section{Results}
\label{sec-2}

The formalism outlined in Section 2 was adopted to self-consistent calculations of $K^-$ nuclear 
quasi-bound states in selected nuclei across the periodic table. Here we present results for the 
P and KM models, supplemented by a phenomenological $K^-$ multinucleon potential $V^{(2)}_{K^-}$ 
determined in the fits of kaonic atom data. We took into account uncertainties of the parameters 
of $V^{(2)}_{K^-}$  shown in Table~\ref{Tab.:ampB}. 
Results for other $K^-N$ interaction models considered in Ref.~\cite{fgNPA16} including more details 
will be discussed elsewhere~\cite{HM17}. 

\begin{figure}[t!]
\begin{center}
\includegraphics[width=0.48\textwidth]{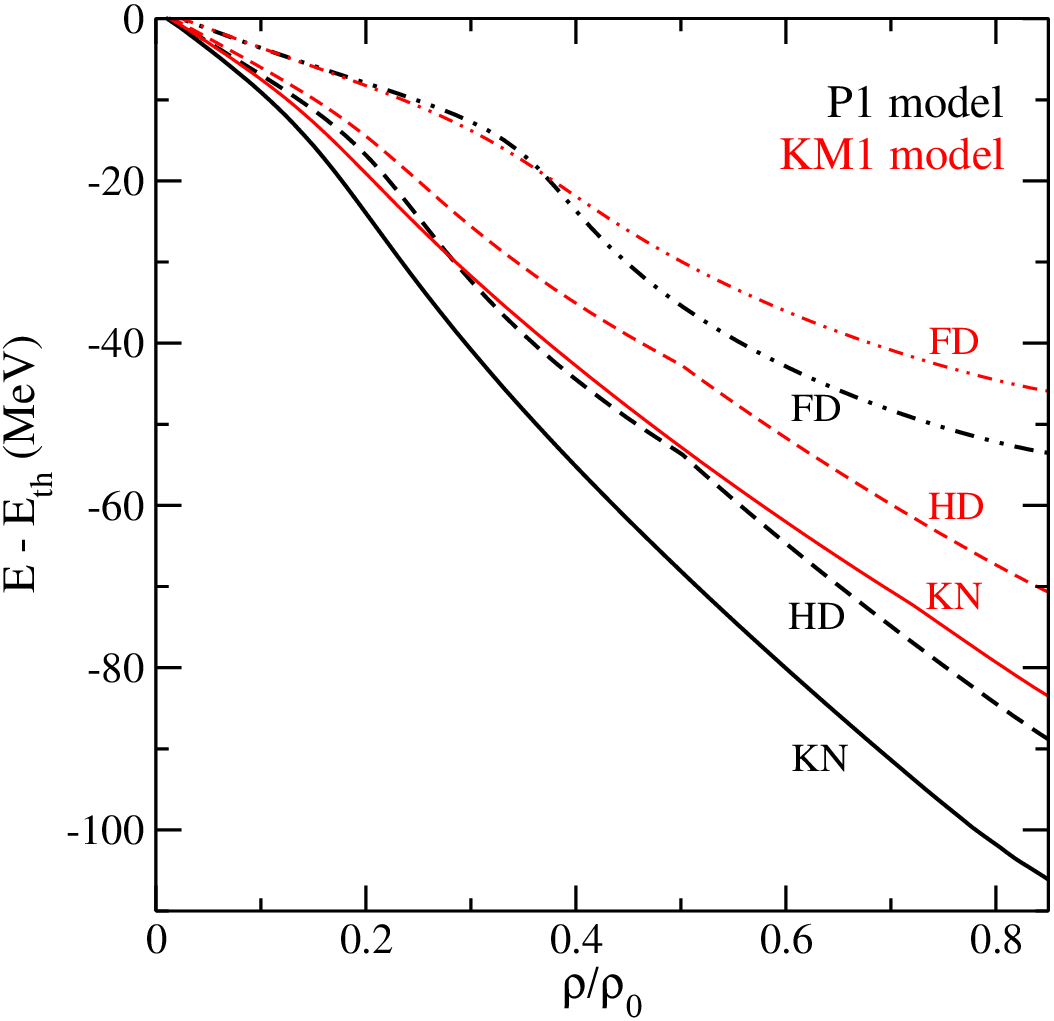} \hspace{5pt}
\includegraphics[width=0.48\textwidth]{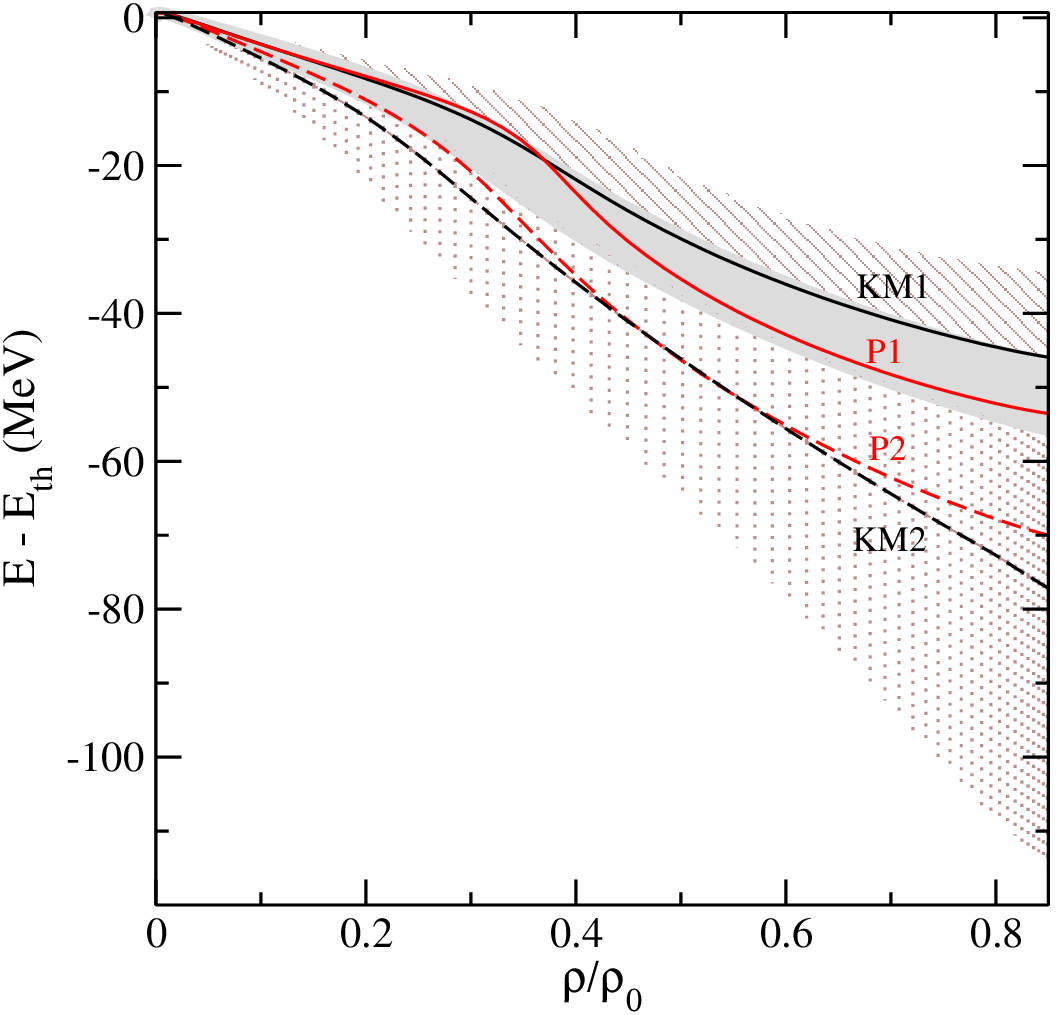}
\end{center}
\caption{
Subthreshold energies probed in the $^{208}$Pb$+K^-$ nucleus 
as a function of relative density $\rho/ \rho_0$, calculated self-consistently in the P1 and KM1 models 
for both options of the $K^-$ multinucleon interaction potential (see text for details) (left) 
compared with the energy shift calculated with the single-nucleon $K^-$ potential (KN, solid 
line). The right panel shows comparison of subthreshold energies probed in considered $K^-N$ amplitude models, supplemented by the FD variant of $V_{K^-}^{(2)}$. The dashed and dotted areas stand for uncertainties and the gray band denotes their overlap.}
\label{rhodelta}
\end{figure}

A characteristic feature of the self-consistently evaluated energy shift $\delta \sqrt{s}$ from~Eq.(\ref{Eq.:deltaEs}) is its strong 
density dependence which plays important role in calculations of kaonic nuclear, as well as atomic states 
using energy dependent chirally motivated $K^-N$ amplitudes.    
The left panel of Fig.~\ref{rhodelta} illustrates the strong density dependence of $\delta \sqrt{s}$ 
in $^{208}$Pb, calculated self-consistently within the P and KM models for $\alpha=1$ (P1 and KM1). 
These models yield for both HD and FD options of $V_{K^-}^{(2)}$ 
smaller energy shift with respect to the $K^- N$ threshold than the 
original single-nucleon potential $V_{K^-}^{(1)}$ (KN). The smallest $\delta\sqrt{s}$ is obtained for 
the full density option FD. 
The P2 and KM2 models (not shown in the figure) yield energy shifts closer to the original KN case. 
The KM2 model gives even slightly larger $\delta \sqrt{s}$ for the HD option than the $K^-$ single-nucleon 
potential due to attractive Re$V_{K^-}^{(2)}$ (positive Re$B$, see Table~\ref{Tab.:ampB}). 
However, the uncertainties for $\alpha=2$ shown in Table~\ref{Tab.:ampB} are so large that the sign of 
Re$B(\alpha=2)$ is insignificant. 
The energy shift for the FD option is in any case shallower than for the original $K^-$ 
single-nucleon potential owing to very strong absorption.

\begin{figure}[t!]
\begin{center}
\includegraphics[width=0.48\textwidth]{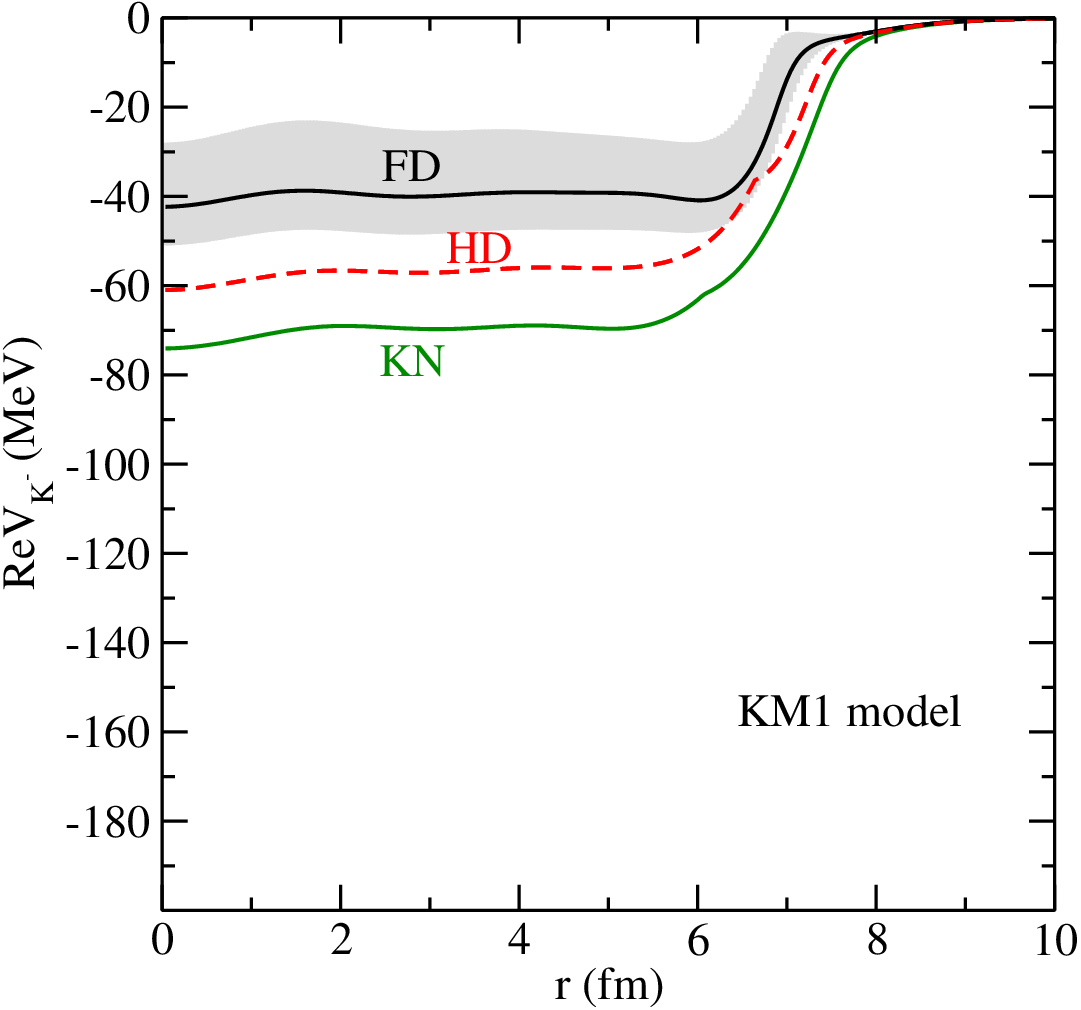} \hspace{5pt}
\includegraphics[width=0.48\textwidth]{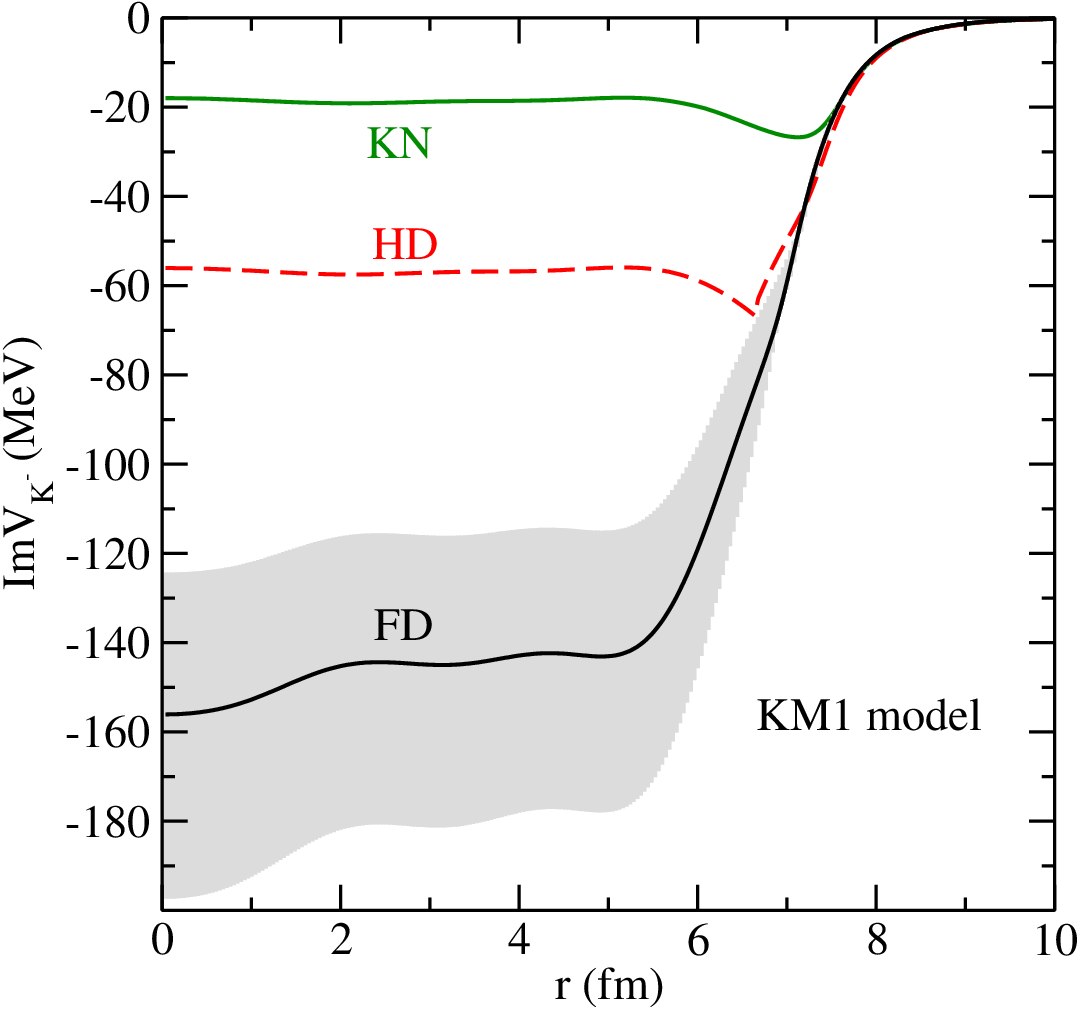}\\[10pt]
\includegraphics[width=0.48\textwidth]{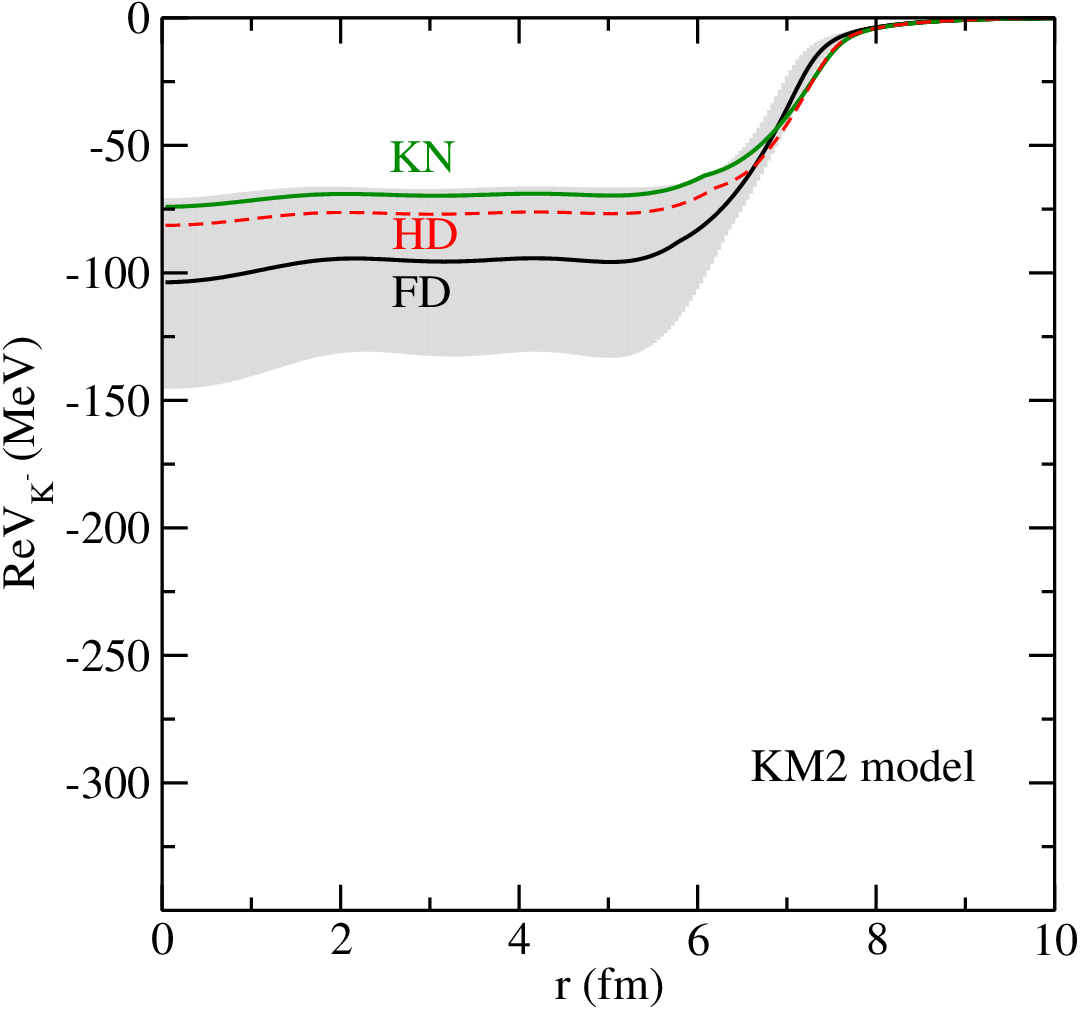} \hspace{5pt}
\includegraphics[width=0.48\textwidth]{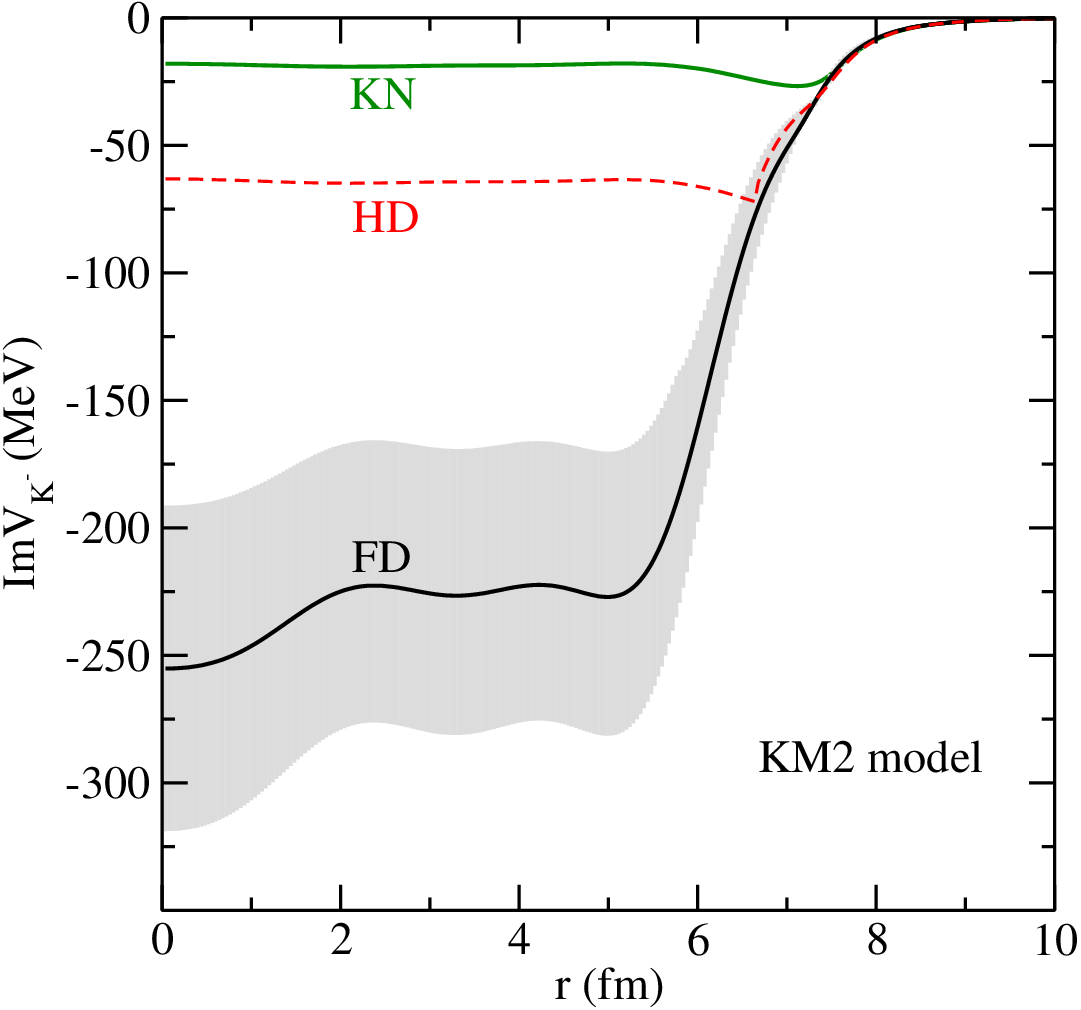}
\end{center}
\caption{The real (left) and imaginary (right) parts of the $K^-$ optical potential in the 
$^{208}$Pb$+K^-$ nucleus, calculated self-consistently in the KM1 (top) and KM2 (bottom) model, 
for two different versions of the $K^-$ multinucleon potential (see text for details). The shaded area stands 
for uncertainties. The single-nucleon $K^-$ potential (KN, green solid line) calculated in the KM model is shown for comparison.}
\label{potknntotal}
\end{figure}

%\begin{figure}[t!]
%\begin{center}
%\includegraphics[width=0.48\textwidth]{ReKM2full+unc.eps} \hspace{5pt}
%\includegraphics[width=0.48\textwidth]{ImKM2full+unc.eps}
%\end{center}
%\caption{The same as in Fig.~2 but for the KM2 model.}
%\label{potknntotal2}
%\end{figure}

In the right panel of Fig.~\ref{rhodelta}, we present subthreshold energies probed by the $K^-$-nuclear 
potential as a function of the nuclear density in $^{208}$Pb, calculated in P and KM interaction models 
with the FD version of the $K^-$ multinucleon potential. The dashed and dotted areas denote uncertainties involved in 
the $K^-$ multinucleon potential $V_{K^-}^{(2)}$  calculated for the KM1 and KM2 models, respectively; the gray band stands for their overlap. The figure 
illustrates the extent of the uncertainties as well as model dependence.   
The energy shifts range from $\approx -35$ to $-115$~MeV in the nuclear center. 
The P1 and KM1 models yield smaller spread in $\delta \sqrt{s}$ due to the uncertainties than the models
with $\alpha=2$. For both values of $\alpha$, the energy shifts calculated using the P and KM models are 
lying within the corresponding uncertainty band, which suggests that the models could be regarded as 
equivalent. 
It is to be noted that in the P2 model, we had to scale the imaginary part of the total $K^-$ potential entering the Klein-Gordon equation by factor 0.8 in order to get numerically stable solution (converged iteration loop). Without the scaling of Im$V_{K^-}$ the energy shift would be smaller than the one presented in the right panel of Fig.~\ref{rhodelta} (for more details see footnote $^1$).
%The energy shift in the P2 model corresponds to $0.8\,{\rm Im}V_{K^-}^{(2)}$~\footnote{In the P2 model, we had to scale the FD option of Im$V_{K^-}^{(2)}$ by factor 0.8 in order to get converged results.}.  
  
%is not presented since we did not obtain fully converged solution.

Fig.~\ref{potknntotal} shows real (left) and imaginary (right) parts of the total $K^-$ potential, calculated 
self-consistently for $^{208}$Pb+$K^-$ in the KM1 (top) and KM2 (bottom) model. The gray shaded 
areas stand for uncertainties in $V_{K^-}^{(2)}$.  
%The real parts of $V_{K^-}$ including multinucleon interactions, both HD and FD options, are shallower than 
%the original single-nucleon $V_{K^-}^{(1)}$ potential since Re$V_{K^-}^{(2)}$ is repulsive. The same holds for the P1 model. 
%In the KM2 model, the contribution from $V_{K^-}^{(2)}$ is attractive since the effective amplitude Re$B$ has a positive sign (see Table~\ref{Tab.:ampB}) and the overall $K^-$ real potential is deeper than the underlying $K^-$ single-nucleon potential. 
%For $\alpha=2$, the uncertainties shown in Table~\ref{Tab.:ampB} are so large that the sign of Re$B(\alpha=2)$, as well as the sign of Re$V_{K^-}^{(2)}$ is insignificant. 
%The larger value of the exponent $\alpha$ in Eq.~\eqref{Vknn}, the larger is sensitivity of extrapolations to the nuclear interior. 
The $K^-$ multinucleon interactions affect the real part of the $K^-$ optical potential markedly less 
than its imaginary part in all considered models, which has crucial consequences for the widths of $K^-$ 
nuclear states.  
The Re$V_{K^-}$ potentials for HD and FD options differ by $\approx 20$~MeV in each interaction model. 
%and the hierarchy of the potentials corresponds to the hierarchy of the energy shifts presented in Fig.~\ref{rhodelta}. 
On the other hand, the imaginary parts of $V_{K^-}$ exhibit much larger dispersion for different versions 
of $V_{K^-}^{(2)}$, as illustrated in Fig.~\ref{potknntotal}, right panels. The $K^-$ multinucleon absorption 
significantly deepens the imaginary part of the $K^-$ optical potential. For the FD option of $V_{K^-}^{(2)}$, the KM model yields $\mid$Im$V_{K^-}\mid \gg$ $\mid$Re$V_{K^-}\mid$ inside the nucleus for both values of $\alpha$, even 
when the uncertainties of the $K^-$ multinucleon potential are taken into account. The same holds for the P model (not shown in the figure).  

\begin{figure}[t!]
\begin{center}
\includegraphics[width=0.48\textwidth]{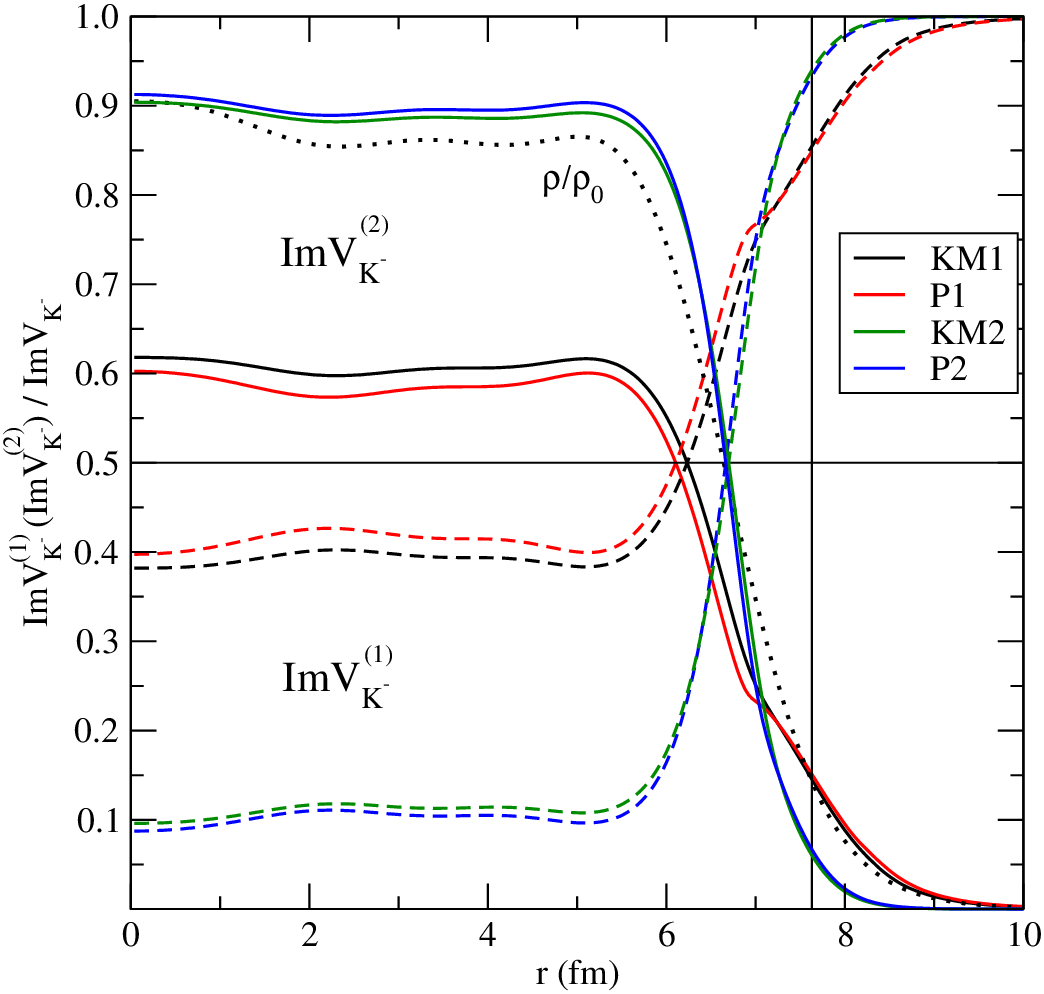}
\end{center}
\caption{The ratio of Im$V_{K^-}^{(1)}$ (dashed line) and Im$V_{K^-}^{(2)}$ (solid line) potentials to the 
total $K^-$ imaginary potential Im$V_{K^-}$ as a function of radius, calculated self-consistently 
for $^{208}$Pb$+K^-$ system  within different meson-baryon interaction models and FD option of the 
$K^-$ multinucleon potential. The relative nuclear density $\rho / \rho_0$ (dotted line) and 
vertical lines denoting $15\%$ of $\rho_0$ are shown for comparison. 
%The shaded areas stand for uncertainties.
}
\label{pomerknn}
\end{figure}

\begin{table}[b!]
\begin{center}
\caption{$1s$ $K^-$ binding energies $B_{K^-}$ and widths $\Gamma_{K^-}$ (in MeV) in various nuclei calculated using the single nucleon 
$K^-N$ amplitudes (denoted KN); plus a phenomenological amplitude $B(\rho/\rho_0)^{\alpha}$, where $\alpha=1$ and 2, 
for 'half-density limit' (HD) and 'full density' option (FD) (see text for details).}
\vspace*{8pt}
 \begin{tabular}{r c|c|c c|c c}
%\hline 
   \multicolumn{3}{c}{KM model}& \multicolumn{2}{|c|}{$\alpha = 1 $}& \multicolumn{2}{|c}{$\alpha = 2$} \\ \hline
   &  & KN & HD & FD &  HD & FD \\ \hline %\hline
 %$^{6}$Li  & $B_{K^-}$ & 24.5 & $\; 10.8$ & not & not & $\; 23.4$ & $\; 18.8$ & not  \\ 
 %     & $\Gamma_{K^-}$ & 45.2 & 115.6 & bound & bound & 121.7  & 160.3 & bound \\ \hline
 %$^{12}$C  & $B_{K^-}$ & 45.3 & $\; 34.4$ & $\; 20.2$ & not & $\; 48.1$ & $\; 44.2$ & not  \\ 
 %     & $\Gamma_{K^-}$ & 43.6 & 113.5 & 181.8 & bound & 124.9  & 191.0 & bound \\ \hline
 $^{16}$O  & $B_{K^-}$ & 45 & $\; 34$ & not & $\; 48$ & not  \\ 
      & $\Gamma_{K^-}$ & 40 & 109 & bound & 121  & bound \\ \hline
 $^{40}$Ca  & $B_{K^-}$ & 59 & $\; 50$ & not & $\; 64$ & not  \\   
      & $\Gamma_{K^-}$ & 37 & 113 & bound & 126  & bound \\ \hline
 %$^{90}$Zr  & $B_{K^-}$ & 68.6 & $\; 56.1$ & $\; 47.0$ & $\; 16.8$ & $\; 71.6$ & $\; 71.2$ & $\; 30.1$ \\   
 %     & $\Gamma_{K^-}$ & 35.7 & 106.8 & 156.0 & 312.3 & 120.4  & 166.9 & 498.6 \\ \hline
 $^{208}$Pb  & $B_{K^-}$ & 78 & $\; 64$ & $\; 33$ & $\; 80$ & 
$\; 53$  \\ 
      & $\Gamma_{K^-}$ & 38 & 108 & 273 & 122  & 429 \\ \hline \hline
%\end{tabular} \label{bkgk1}
%\end{center}
%\end{table} 
%\begin{table}[h!]
%\begin{center}
%\caption{The same as in Table~2 but for the P model.}
%
%\vspace*{8pt}
% \begin{tabular}{r c|c|c c|c c}
%%\hline 
   \multicolumn{3}{c}{P model}& \multicolumn{2}{|c|}{$\alpha = 1 $}& \multicolumn{2}{|c}{$\alpha = 2$} \\ \hline
%   &  & KN & HD & FD &  HD & FD \\ \hline %\hline
 %$^{6}$Li  & $B_{K^-}$ & 37.5 & $\; 21.3$ & not & not & $\; 35.9 $ & $\; 28.2 $ & not  \\  
 %     & $\Gamma_{K^-}$ & 39.6 & 112.3 & bound & bound & 132.8  & 183.4 & bound \\ \hline
 %$^{12}$C  & $B_{K^-}$ & 63.6 & 49.5 & $\; 34.6$ & not & $\; 63.5 $ & $\; 56.6 $ & not  \\  
 %     & $\Gamma_{K^-}$ & 28.4 & 96.1 & 165.0 & bound & 122.3  & 196.3 & bound \\ \hline
 $^{16}$O  & $B_{K^-}$ & 64 & 49 & not & $\; 63$ & not  \\  
      & $\Gamma_{K^-}$ & 25 & 94 & bound & 117  & bound \\ \hline
 $^{40}$Ca  & $B_{K^-}$ & 81 & 67 & not & $\; 82$ & not  \\  
      & $\Gamma_{K^-}$ & 14 & 95 & bound & 120  & bound \\ \hline
 %$^{90}$Zr  & $B_{K^-}$ & 90.0 & 74.1 & $\; 62.3$ & $\; 18.9$ & $\; 87.1$ & $\; 84.7$ & not  \\  
 %     & $\Gamma_{K^-}$ & 12.4 & 87.8 & 135.5 & 339.6 & 113.7  & 163.8 & bound \\ \hline
 $^{208}$Pb  & $B_{K^-}$ & 99 & 82 & $\; 36$ & $\; 96$ & $\; 47^{*}$  \\ 
      & $\Gamma_{K^-}$ & 14 & 92 & 302 & 117  & $412^{*}$ \\
\hline
\end{tabular}\label{bkgk2}
\end{center}
\vspace*{5pt}
\hspace*{25pt} $^*$the solution of Eq. \eqref{KG} for Im$V_{K^-}$ scaled by factor 0.8
\end{table} 
The particular role of $K^-$ single- and multinucleon absorptions with respect to the nuclear density is 
illustrated in Fig.~\ref{pomerknn}. Here we compare individual contributions of $K^-$ single-nucleon and 
multinucleon absorptions to the total $K^-$ absorption, expressed as a fraction of Im$V_{K^-}^{(1)}$ and 
Im$V_{K^-}^{(2)}$ with respect to the total imaginary $K^-$ potential Im$V_{K^-}$, 
calculated self-consistently for $^{208}$Pb+$K^-$ in the P and KM models. The density $\rho / \rho_0$ 
(thin dotted line) is shown for comparison. %The shaded areas denote uncertainties in $V^{(2)}_{K^-}$. 
The relative contribution of Im$V_{K^-}^{(1)}$ and Im$V_{K^-}^{(2)}$ to $K^-$ 
absorption is changing with radius (density) because of the different range of corresponding potentials. At 
the nuclear surface, the $K^-$ absorption on a single nucleon dominates, while it is reduced in the nuclear 
interior due to vicinity of $\pi \Sigma$ threshold and the multinucleon absorption prevails. 
The single-nucleon $K^-N$ absorption in the nuclear medium is more suppressed in the models with $\alpha=2$ 
since the self-consistent value of $\sqrt{s}$ at $\rho_0$ is closer to the $K^-N \rightarrow \pi \Sigma$ 
threshold than in the models with $\alpha=1$. 
The analysis of Friedman and Gal~\cite{fgNPA16} showed that the fractions 
of $K^-$ absorption on a single nucleon ($\sim 75\%$) and several nucleons ($\sim 25\%$) from the bubble 
chamber experiments are sensitive to about $15\%$ of nuclear density (denoted in Fig.~\ref{pomerknn} by 
vertical black line). At this density, the ratios ${\rm Im}V_{K^-}^{(2)} / {\rm Im}V_{K^-}^{(1)}$ 
are lower than experimental fractions of $K^-$ absorption at rest \cite{bubble1, bubble2, bubble3} due to 
different self-consistent values of $\delta\sqrt{s}$ for kaonic and nuclear states. 
However, we stress that one has to compare corresponding widths, rather than Im$V_{K^-}^{(1)}$ and 
Im$V_{K^-}^{(2)}$, for proper confrontation with experiment.

In Table~\ref{bkgk2} we present $1s$ $K^-$ binding energies $B_{K^-}$ and widths $\Gamma_{K^-}$, calculated 
in the KM and P models. The $K^-$ binding energies and widths calculated only   
with the underlying $K^-$ single-nucleon potential are shown for comparison. When $K^-$ multinucleon interactions are included, the $K^-$ widths increase considerably. 
For the HD option the $K^-$ widths are of order $\sim 100$~MeV and exceed significantly the corresponding 
$K^-$ binding energies.
  
%The $K^-$ binding energies for HD option change only slightly --- they decrease in KM1, P1, and P2 models and increase in KM2 model. The binding energies are much smaller than the corresponding $K^-$ widths which are of order $\sim 100$~MeV.
For the $K^-$ interaction models with the FD multinucleon potentials $V_{K^-}^{(2)}$, the antikaon is unbound in the majority of nuclei.  
We found $1s$ $K^-$ quasi-bound states in $^{208}$Pb, however, the $K^-$ widths are huge, one order of magnitude larger than the binding energies \footnote{In the case of the P2 model, we present the solution of the Klein-Gordon equation~\eqref{KG} for $^{208}$Pb with Im$V_{K^-}$ scaled by factor 0.8 since the calculation with the full imaginary potential is not numerically stable due to extremely strong $K^-$ absorption --- the non-converged $\Gamma_{K^-} > 500$~MeV while the corresponding $B_{K^-} < 15$~MeV.}. These conclusions remain valid even when the uncertainties in the 
$K^-$ multinucleon potential are taken into account.  

%\newpage

\section{Conclusions}
\label{sec-3}
This work reports on calculations of $K^-$ nuclear quasi-bound states performed 
using a $K^-$ single-nucleon potential derived within two chirally motivated meson-baryon 
coupled-channel models P and KM, supplemented by a phenomenological potential representing the $K^-$ multinucleon interactions. Parameters of the phenomenological potential were recently fitted by Friedman and Gal \cite{fgNPA16} to kaonic atom data for each meson-baryon interaction model separately. Moreover, in the analysis of Ref.~\cite{fgNPA16} the single-nucleon $K^-$ potential constructed within the P and KM chiral models together with a phenomenological $K^-$ multinucleon potential $V_{K^-}^{(2)}$ was confronted with the branching ratios of $K^-$ single-nucleon absorption at rest for the first time. The fractions of $K^-$ single-nucleon absorption calculated within these two models are in agreement with the data from bubble chamber experiments.

Since the kaonic-atom data probe the $K^-$ optical potential reliably up to at most $\sim 50\%$ of $\rho_0$, 
two scenarios for extrapolating $V_{K^-}^{(2)}$ to higher densities $\rho \geq 0.5\rho _0$ were considered. 
Moreover, uncertainties of the parameters of the phenomenological $K^-$ multinucleon 
potentials were taken into account in order to verify that the results are sufficiently robust. 

The fractions of $K^-$ single-nucleon and multinucleon absorption in the medium were evaluated. 
At the surface of a nucleus, the fractions are in accordance with experimental data. 
In the nuclear interior, the $K^-$ single-nucleon absorption is reduced due to the vicinity of $\pi \Sigma$ threshold and the $K^-$ multinucleon absorption prevails. 

The $K^-$ multinucleon interactions were found to cause radical increase of $K^-$ widths. 
In vast majority of nuclei the widths exceed considerably $K^-$ binding energies. The FD variant of the phenomenological 
potential does not even yield any $K^-$ bound states in most of the nuclei. Calculations performed for 
other nuclei 
and other recent $K^-N$ interaction models considered in Ref.~\cite{fgNPA16} confirmed our conclusions 
concerning the decisive effect of $K^-$ multinucleon interactions on $K^-$ widths in nuclei~\cite{HM17}. In view of our results, it would be desirable to explore the role of the $K^-$ multinucleon processes in few-body systems as well.

The main message of the present study is that the $K^-$-nuclear quasi-bound states in many-body nuclear systems, if they ever exist, have huge widths, considerably exceeding $K^-$ binding energies. Their identification in experiment thus seems impossible.

\section*{Acknowledgements}
We wish to thank E. Friedman and A. Gal for valuable discussions, and A. Ciepl\'{y} 
and M. Mai for providing us with the free $K^-N$ scattering amplitudes.
This work was supported by the GACR Grant No. P203/15/04301s.\\  
J. Hrt\'{a}nkov\'{a} acknowledges financial support from the CTU-SGS Grant No. SGS16/243/OHK4/3T/14.\\  
Both J. H. and J. M. acknowledge the hospitality extended to them at the Racah Institute of Physics, 
The Hebrew University of Jerusalem, during their collaboration visit in November 2016. 
J. M. acknowledges financial support of his visit provided by the Racah Institute of Physics. 
J. H. acknowledges financial support of the Czech Academy of Sciences which enabled her stay at the Hebrew University.

\end{document}